\documentstyle [preprint,pra,aps]{revtex}
\begin {document}
\draft
\preprint {pra3.0}
\title {
Geometry of one-dimensional wave propagation
}
\author {M. Kitano}
\address {
Department of Electronics, Kyoto University
}
\date {\today }
\maketitle
\begin {abstract}
We investigate the geometrical features of one-dimensional
wave propagation,
whose dynamics is described by the (2+1)-dimensional Lorentz group.
We find  many interesting geometrical ingredients such as
spinorlike behavior of
wave amplitudes, gauge transformations, Bloch-type equations,
and Lorentz-group Berry phases.
We also propose an optical experiment to verify these effects.
\end {abstract}


\narrowtext
\section {Introduction}
\label {sec:introduction}
The Berry phases \cite {berry} produced by rotations have
been most extensively
studied for various physical systems.
Another interesting class consists of Berry phases produced
by Lorentz transformations.
Chiao and Jordan \cite {chiao-jordan} showed that
Lorentz-group Berry phases
could be observed
as a change of phase of the electromagnetic field
in squeezed states.
Successive squeezing operations for either light or microwaves
by degenerate parametric amplifiers
induce a phase shift that can be interpreted geometrically.
These kinds of experiments have drawn considerable interest because
they can be viewed as the quantum optical realization of
the Lorentz kinematics.\cite {yurke,han}

Another optical manifestation of the Lorentz-group Berry's phase
was proposed and experimentally verified \cite {kitano-yabuzaki}.
It is shown that when one cycles light through a sequence
of partially polarized states, the light acquires a geometrical phase.

It is well known that the
wave propagation in one dimension (1D) can be
described by the Lorentz-group dynamics,
therefore, one can expect Lorentz-group Berry phases in wave systems.
In this paper, we present Berry phases and related
geometrical features seen in the 1D wave propagation in
inhomogeneous media.

In Sec.\ \ref {sec:1dwave}, from the 1D wave equation
with quasi-periodic potential we derive
a Schr\"{o}dinger-like equation, which explicitly exhibits
the Lorentz dynamics via the generators of SL(2,R).
The dynamical variables of the equation are
the two quadrature wave amplitudes, which
form a two-column state vector.

In Sec.\ \ref {sec:blochreal}, a SL(2,R)-Bloch equation
is derived from the Schr\"{o}dinger-like
equation.
The equation is very similar to the conventional SU(2)-Bloch
equation for the spin-1/2
except that the SL(2,R)-Bloch vector moves on a conical surface
not on a sphere.
The similarity could be very useful for the geometrical understanding of
Lorentz dynamics,
especially because the SU(2)-Bloch model has been successfully employed
in various fields such as nuclear magnetic resonances,
nonlinear optics, polarization optics, interferometry,
and atomic physics, etc.

In Sec.\ \ref {sec:twovalue}, we discuss two-valuedness of the state vectors.
A Bloch vector represents a pair of state vectors with opposite signs.
When a Bloch vector circles around the apex of a cone, the corresponding
state vector changes its sign.
It is shown that the two-valuedness comes from a symmetry
of the wave equation with respect to the translation of
half wavelength.

Extending the above geometrical situation,
we introduce in Sec.\ \ref {sec:gauge} a local gauge transformation
associated with
the deformation of space coordinate.

So far, we have assumed that the state vector is real
because the potential function in the wave equation is real.
It turns out, however, the extension to complex wave functions
is very convenient, because a complex solution could represent
two linearly independent real solutions at once.
In Sec.\ \ref {sec:blochcomplex}, we derive a Bloch equation for the case of
complex amplitudes.
In this case,
a Bloch vector  corresponds to a set of state vectors that differ only
in phase factors.
The Bloch vector moves on the surface of
hyperboloid.
This is the arena where the Berry phase plays a role.
In Sec.\ \ref {sec:geometricalphase}, we define the Berry phase for
a cyclic evolution of the system.

Finally, we propose a feasible optical experiment where we can observe
the Berry phase for 1D wave propagation.

%====================================
\section {1D Wave Equation and SL(2,R)}
%====================================
\label {sec:1dwave}

Let us start with a one-dimensional wave equation
(the Helmholtz equation)
for a monochromatic wave $\Psi (x){\rm e}^{-{\rm i}\omega  t}$
\begin {equation}
 \left \{\frac {{\rm d}^2}{{\rm d} x^2} + k^2[1+\epsilon (x)]\right \}
\Psi (x)=0,
 \label {eq:1}
\end {equation}
where $k=k(\omega )$ is the wave number in free ($\epsilon =0$) space.

We assume the spatial variation of the refractive index or the potential
$\epsilon (x)$
has the following form:
\begin {equation}
\epsilon (x)=\epsilon _0(x)
+2\epsilon _{\rm c}(x)\mathop {\rm cos}\nolimits
2kx+2\epsilon _{\rm s}(x)\mathop {\rm sin}\nolimits  2kx.
\label {eq:2}
\end {equation}
The spatial modulation at wave number $2k$ strongly modifies the
propagation of the wave with wave number $k$ owing to the Bragg effect.
The DC component $\epsilon _0$ also has substantial influences
through the Bragg resonance condition.

We represent the wave $\Psi (x)$ with mean wave number $k$
as
\begin {equation}
\Psi (x)=A(x)\mathop {\rm cos}\nolimits  kx +
B(x)\mathop {\rm sin}\nolimits  kx,
\label {eq:3}
\end {equation}
where $A(x)$ and $B(x)$ are slowly varying envelope functions.
These envelopes are normally considered to be real functions,
since $\epsilon (x)$ in Eq.\ (\ref {eq:1}) is real.
However, the extension to complex functions is sometimes very useful as
will be discussed in Sec.\ \ref {sec:blochcomplex}.

Substitution of Eqs.\ (\ref {eq:2}) and (\ref {eq:3}) into
Eq.\ (\ref {eq:1}) gives
\begin {equation}
  \frac {{\rm d}}{{\rm d} x}\left [\begin {array}{c}A\\B\end {array}\right ]=
  \frac {k}{2}\left [\begin {array}{cc}
   \epsilon _{\rm s} & -\epsilon _{\rm c}+\epsilon _0 \\
   -\epsilon _{\rm c}-\epsilon _0 & -\epsilon _{\rm s}
   \end {array}\right ]
   \left [\begin {array}{c}A\\B\end {array}\right ],
\label {eq:4}
\end {equation}
which can be rewritten in a
Schr\"{o}dinger-like form
\begin {equation}
{\rm i} k^{-1}\frac {{\rm d}}{{\rm d} x}|\psi \rangle
= - \bbox {\epsilon }(x)\cdot \bbox {K}|\psi \rangle .
\label {eq:5}
\end {equation}
with the column vector
$
|\psi \rangle =(A,B)^{\rm  T}.
$
Here we introduced
$
\bbox {\epsilon }(x)=(-\epsilon _{\rm s},\epsilon _{\rm c},\epsilon _0)
= (\epsilon _1, \epsilon _2, \epsilon _3),
$
and
$
\bbox {K} = (K_1, K_2, J_3),
$
with
\begin {equation}
  K_1 = {\rm i} \sigma _3/2,\;
  K_2 = {\rm i} \sigma _1/2,\;
  J_3 = \sigma _2/2,
  \label {eq:6}
\end {equation}
where $\sigma _i (i=1,2,3)$ are the Pauli's spin matrices.

We have assumed that $A(x)$ and $B(x)$ do not change appreciably over
distances of the order of $k^{-1}$
(slowly-varying-envelope approximation)
and neglected the terms with wavelength $\pm  3k$
(secular-term approximation).

The matrices (\ref {eq:6}) are the generators of
SL(2,R), the two dimensional real unimodular group,
and satisfy the commutation relations
\begin {equation}
  [K_1,K_2]=-{\rm i} J_3,\hskip 1em\relax
[K_2, J_3]={\rm i} K_1, \hskip 1em\relax  [J_3, K_1]={\rm i} K_2.
\label {eq:7}
\end {equation}
The $(A, B)$ plane is squeezed in one direction and stretched
in the orthogonal direction by the generators $K_1$ or $K_2$,
and rotated by $J_3$.

The group SL(2,R) has the close connection
(locally isomorphic) to the (2+1)-dimensional
Lorentz group SO(2,1) as does the group SU(1,1).
Instead of Eq.\ (\ref {eq:3}), if we use a representation
\begin {equation}
  \Psi (x) = {\cal A}(x){\rm e}^{-{\rm i} kx} +
{\cal B}(x){\rm e}^{{\rm i} kx},
\label {eq:8}
\end {equation}
then we have the SU(1,1)-Schr\"{o}dinger equation with
  $K'_1 = {\rm i}\sigma _1/2$,
  $K'_2 = -{\rm i}\sigma _2/2$,
  $J'_3 = \sigma _3/2$,
which satisfy the same commutation relations as Eq.\ (\ref {eq:7}).
%(more on SU(1,1))
In the following discussion, we mostly use SL(2,R) rather than
SU(1,1).
%======================================================
\section {SL(2,R)-Bloch Equation -- Real Amplitude Case}
%======================================================
\label {sec:blochreal}

Equation (\ref {eq:5}) has a form similar to the Schr\"{o}dinger
equation for the spin-1/2 ($\bbox {J}$) in time varying magnetic
fields $\bbox {B}(t)$;
${\rm i}\hbar ({\rm d}/{\rm d} t)|\psi \rangle  =
-\bbox {B}(t)\cdot \bbox {J}|\psi \rangle $.
The underlying group for the spin-1/2 dynamics is
SU(2), which is the covering group of the three dimensional
rotation group, SO(3).
The Bloch equation derived from the spin-1/2 Schr\"{o}dinger
equation apparently reflects the structure of those groups;
the Bloch vector moves on a sphere, and
the Berry phase for spin-1/2 can be related to the surface area of
the sphere.

In order to see the geometrical structure of
SL(2,R)-dynamics, let us derive the corresponding Bloch equation.
(A more detailed derivation will be given in Sec.\ \ref {sec:blochcomplex}.
See also Ref.\ \cite {dattoli}.)

First we define the $2\times 2$ density matrix $\rho $ as
\begin {equation}
  \rho  = 2{\rm i} |\psi \rangle \langle \psi |J_3={\rm i}
\bbox {s}\cdot \bbox {K},
\label {eq:9}
\end {equation}
where $\langle \psi |$ represents the row vector $(A, B)$, and
$\bbox {s}=(s_1, s_2, s_3)$ the Bloch vector.
We note,
\begin {equation}
  \bbox {s} = (2AB, -A^2+B^2, A^2+B^2).
\label {eq:13}
\end {equation}

The equation of motion for $\rho $ can be obtained from Eq.\ (\ref {eq:5}) as
\begin {equation}
  {\rm i} k^{-1}\frac {{\rm d}\rho }{{\rm d} x} =
   [-\bbox {\epsilon }\cdot \bbox {K}, \rho ].
\label {eq:10}
\end {equation}
Using the commutation relations (\ref {eq:7}), we have the equation of
motion for $\bbox {s}$
\begin {equation}
  k^{-1}\frac {{\rm d}\bbox {s}}{{\rm d} t}=\bbox {\epsilon }
\stackrel {\sim }{\times }\bbox {s},
\label {eq:11}
\end {equation}
where the SU(1,1) vector product defined
as
\begin {equation}
  \bbox {a}\stackrel {\sim }{\times }\bbox {b}
  =(a_2b_3-a_3b_2, a_3b_1-a_1b_3, -a_1b_2+a_2b_1).
\label {eq:12}
\end {equation}
for two vectors,
$\bbox {a}=(a_1, a_2, a_3)$ and
$\bbox {b}=(b_1, b_2, b_3)$.
We also define the SU(1,1) scalar product
\begin {equation}
  \bbox {a}\stackrel {\sim }{\cdot }\bbox {b} = a_1b_1+a_2b_2-a_3b_3.
\label {eq:14}
\end {equation}
Then from Eq.\ (\ref {eq:13}), we have
\begin {equation}
  \bbox {s}\stackrel {\sim }{\cdot }\bbox {s} = 0,
\hskip 1em\relax  s_3 \ge  0,
\label {eq:15}
\end {equation}
which means the Bloch vector $\bbox {s}$ moves on the surface of
the (upper) cone depicted in Fig.\ \ref {fig1}.

When $\bbox {\epsilon }(x)$ is constant,
trajectories for $\bbox {s}$ are very simple;
the vector $\bbox {s}$ lies in a plane that is orthogonal to the
vector $\mathaccent "707E {\bbox {\epsilon }}=
(\epsilon _1,\epsilon _2,-\epsilon _3)$,
because
\begin {equation}
  \frac {{\rm d}\bbox {s}}{{\rm d} x}\stackrel
{\sim }{\cdot } \bbox {\epsilon } =
  \frac {{\rm d}\bbox {s}}{{\rm d} x}\cdot  \mathaccent "707E
{\bbox {\epsilon }} = 0.
\label {eq:39}
\end {equation}

A trajectory, which is the intersection
of a plane and the cone
could be an ellipse, a parabola, or a hyperbola
according as $\epsilon _3^2$ is greater than, equal to, or less than
$\epsilon _1^2+\epsilon _2^2$.
The elliptical, bound trajectories correspond to the propagating waves in
the conduction bands and the hyperbolic, unbound
trajectories correspond to
the evanescent waves in the forbidden bands.

Owing to the geometrical simplicity and the similarity to the
conventional Bloch equation, the SL(2,R)-Bloch equation could be very useful.

%=======================
\section {Two-valuedness}
%=======================
\label {sec:twovalue}

In this section, we study a subtler geometrical
feature of SL(2,R)-Bloch equation.
We note the correspondence between $(A, B)$ and $\bbox {s}$
given by Eq.\ (\ref {eq:13}) is not one to one but two to one;
a single vector $\bbox {s}$ represents two state vectors
with opposite signs: $(A, B)$ and $(-A,-B)$.

At first sight, these two state vectors may seem to represent
two distinct physical situations.
Within the framework of the present approximation (slowly
varying envelopes), however, the two situations are
virtually identical.
Figure \ref {fig2} represents the local pictures of waves
with opposite polarities (real lines).
A potential with period $\sim \pi k^{-1}$ is also shown (dashed lines).
We note the dispositions of the waves relative to the potential
are the same and, therefore, the waves would evolve identically.

This degeneracy comes from the fact that
the potential (\ref {eq:2}) is invariant under
the translation $x \rightarrow  x + n\pi  k^{-1}$ ($n$: integer),
as far as the variation of $\bbox {\epsilon }(x)$ over
the distance $n\pi  k^{-1}$ can be neglected;
\begin {equation}
  \epsilon (x+n\pi  k^{-1})\sim \epsilon (x).
\label {eq:16}
\end {equation}
On the other hand, from Eq.\ (\ref {eq:3}),
we see that the translation induces
$(A, B) \rightarrow  (-1)^n(A, B)$.

In order to see the geometrical structure of the two-valued representation,
we can make a cone from the $(A,B)$ plane as shown in Fig.\ \ref {fig3}.
The $(A,B)$ plane is cut along a line from the origin
($+A$ axis in this example)
and is wrapped to make a twofold cone with apex angle of 60 degrees.
When the vector $\bbox {s}$ on the cone encircles a closed curve
around the apex, we see that
the corresponding $(A, B)$ gains a factor $(-1)^n$, where $n$ is
the winding number of the curve.

Even though, two vectors $(A,B)$ and $(-A,-B)$ represent
(almost) the same physical situation,
we can conveniently detect the minus sign by interferometric methods,
as that for spin 1/2 rotated by $2\pi$ \cite{rauch-werner}.
(The two cases have good analogy, but it is only for the latter case
that interference experiments are absolutely
required and even in principle there is no other way to detect
the minus sign.)

Figure \ref {fig4} shows an example of interferometry.
We have two trajectories both of which start from $(A,B)=(1,0)$
at $x=0$.
One evolves with a constant $\bbox {\epsilon }$ and
the other with $-\bbox {\epsilon }$.
After some evolution, they reach conjugate points in the $(A,B)$ plane.
Looking at the corresponding wave forms [Fig.\ \ref {fig4}(b)],
we notice that the phase shifts accumulate to make signs opposite.

This situation reminds us of
the sign change of a spinor rotated by $2\pi $,
the Aharonov-Bohm effect with an infinite solenoid,
the sign change around the degeneracy of the eigenstates of
real-Hamiltonian systems \cite {longuet-higgins,berry-wilkinson}, and
the (relative) configuration space of two identical particles
in two-dimensional space \cite {leinaas-myrheim}.
In terms of the second analogy, the topological
magnetic flux of $\pi \hbar  c/e$ is required
to account for the sign change or the phase shift of $\pi $.
The amount of flux is consistent with
the (singular) curvature at the cone apex or the apex angle.

Mathematically speaking, SL(2,R) is a two-valued representation of
the (2+1)-dimensional Lorentz group, SO(2,1), as SU(2) is for the
three-dimensional rotation group, SO(3).
It is very interesting that the two-valued representation which is believed
peculiar to the quantum regime manifests itself in the classical context.

It should be noted that the cone introduced here and that
for the Bloch vector discussed in the previous section have
different apex angles and different parametrization:
\begin {equation}
\overline {\bbox {s}} = (s_1, s_2, \sqrt {3}s_3)/2\sqrt {s_3},
\label {eq:17}
\end {equation}
where $\overline {\bbox {s}}$ is a vector in Fig.\ \ref {fig3}.

%=========================================================
\section {Gauge Transformation}
%=========================================================
\label {sec:gauge}

In order to generalize the discussion in the previous section,
we introduce a (local) gauge transformation.
We conveniently
use the SU(1,1) amplitudes [see Eq.\ (\ref {eq:8})] ${\cal A}$ and ${\cal E}$:
\begin {equation}
{\cal A}=(A+{\rm i} B)/2,\hskip 1em\relax  {\cal E}=
\epsilon _{\rm c}+{\rm i}\epsilon _{\rm s}.
\label {eq:18}
\end {equation}
With this notation, the Schr\"{o}dinger equation (\ref {eq:5})
can be represented in a scalar form as
\begin {equation}
2{\rm i} k^{-1}\frac {{\rm d}}{{\rm d} x}{\cal A}={\cal E}{\cal A}^*
+\epsilon _0{\cal A},
  \label {eq:19}
\end {equation}
where ${\cal A}^*$ stands for the complex conjugate of
${\cal A}$.

Here we consider a space coordinate $x'$ which is slightly
deviated from $x$ as
\begin {equation}
x' = x - \xi (x).
\label {eq:20}
\end {equation}
We assume the deviation $\xi $ is small enough to assure
${\cal A}(x-\xi )\sim {\cal A}(x)$, ${\cal E}(x-\xi )\sim {\cal E}(x)$,
and $\xi (x-\xi )\sim \xi (x)$ for any $x$.
In the $x'$ coordinate, ${\cal A}'$ and ${\cal E}'$ are defined as
\begin {equation}
{\cal A}' = {\cal A}{\rm e}^{-{\rm i} k\xi }, \hskip 1em\relax
{\cal E}' = {\cal E}{\rm e}^{-2{\rm i} k\xi }
  \label {eq:21}
\end {equation}
respectively, and the equation of motion transforms as
\begin {equation}
2{\rm i} k^{-1}\frac {{\rm d}}{{\rm d} x'}{\cal A}'={\cal E}'{\cal A}'^*
+\left [\epsilon _0+2\frac {{\rm d} \xi }{{\rm d} x'}\right ]{\cal A}'.
\label {eq:22}
\end {equation}
Here we have an extra term $2{\rm d}\xi /{\rm d} x'$ which accounts for
the gauge transformation.
This term can be interpreted as a gauge field.

In Eqs.\ (\ref {eq:2}) and (\ref {eq:3}),
in order to define $\epsilon $, $A$, and $B$, we needed
a long yardstick
accurately graduated in $2\pi  k^{-1}$.
However, with the help of the above equations, we can compare
the wave amplitudes for two observers whose rulers are not
necessarily accurate in a long span.

We note that the discussion in the previous section can be
reproduced by considering the case of constant deviation
$\xi (x)=n\pi  k^{-1}$,
which yields ${\cal A}'=-{\cal A}$, ${\cal E}'={\cal E}$,
and ${\rm d}\xi /{\rm d} x'=0$.
We consider two coordinate systems $x$ and $x'$ shifted by
$n\pi  k^{-1}$, and
a third coordinate system $x'' = x + \xi (x)$,
with $\xi (x) = 0$ for $x < x_1$ and
$\xi (x) = n\pi  k^{-1}$ for $x > x_2$;
in the interval of $x_1 < x < x_2$, $x''$ smoothly connects
$x$ and $x'$.
The integration of the gauge field
\begin {equation}
l=\intop \nolimits _{x_1}^{x_2}2\frac {{\rm d}\xi (x)}{{\rm d} x}{\rm d} x =
2n\pi  k^{-1},
 \label {eq:23}
\end {equation}
which is independent of the local behavior of $\xi $,
amount to the sign factor $\exp (-{\rm i} nkl/2)=(-1)^{n}$ between
$(A, B)$ and $(A', B')$.

%=========================================================
\section {SL(2,R) Bloch Equation --- Complex Amplitude Case}
%=========================================================
\label {sec:blochcomplex}

{}From now on let us consider
$|\psi \rangle =(\alpha ,\beta )^{\rm  T}$ as complex amplitudes.
Before proceeding, physical meaning of complex amplitudes should be
clarified.
At first sight of Eq.\ (\ref {eq:1}), which has real coefficients, no complex
amplitudes seems required.
In fact, if we decompose $|\psi \rangle $ into real and imaginary parts
\begin {equation}
|\psi \rangle =|\psi '\rangle + {\rm i}|\psi ''\rangle ,
\end {equation}
then each of $|\psi '\rangle $ and $|\psi ''\rangle $ is a real solution to
Eq.\ (\ref {eq:5}).
Therefore, a complex solution could represent
two independent real solutions.
The independence is assured by
$\alpha '\beta ''-\alpha ''\beta ' \not = 0$,
where $|\psi '\rangle =(\alpha ',\beta ')^{\rm  T}$ and
$|\psi '\rangle =(\alpha '',\beta '')^{\rm  T}$.
A complex solution is convenient in the sense that
any real solutions can be represented as
${\rm  Re}(z|\psi \rangle )$ with a complex number $z$.

Starting from the SL(2, R)-Schr\"{o}dinger equation (\ref {eq:5})
with complex $|\psi \rangle $:
\begin {eqnarray}
  {\rm i} k^{-1}\frac {{\rm d}}{{\rm d} x}|\psi \rangle =N(x)|\psi \rangle ,
\label {eq:25}\\
  N(x) = \epsilon _1 K_1 +\epsilon _2 K_2 +\epsilon _3 J_3,
\label {eq:26}
\end {eqnarray}
let us derive the corresponding Bloch-like equation.
The adjoint equation\cite {garrison} is
\begin {eqnarray}
  {\rm i} k^{-1}\frac {{\rm d}}{{\rm d} x}|\varphi \rangle =N^{\dagger
}(x)|\varphi \rangle ,
\label {eq:27}\\
  N^{\dagger }(x) = -\epsilon _1 K_1 -\epsilon _2 K_2 +\epsilon _3 J_3,
\label {eq:28}
\end {eqnarray}
or equivalently,
\begin {equation}
  -{\rm i} k^{-1}\frac {{\rm d}}{{\rm d} x}\langle \varphi |
=\langle \varphi |N(x).
\label {eq:29}
\end {equation}

In our case, $N$ and $N^{\dagger }$ are related via
\begin {equation}
  N^{\dagger } = R N R^{-1},
\label {eq:30}
\end {equation}
where $R = -R^{-1} = \exp ({\rm i} \pi  J_3) = 2{\rm i} J_3$.
With use of this relation, we can rewrite the adjoint equation (\ref {eq:27})
as
\begin {equation}
  {\rm i} k^{-1}\frac {{\rm d}}{{\rm d} x}R|\varphi \rangle =
N(x)R|\varphi \rangle .
\label {eq:31}
\end {equation}
and find that $R|\varphi \rangle $ and $|\psi \rangle $ obey the same equation.
Hence, if $R|\varphi (0)\rangle =|\psi (0)\rangle $, then
$R|\varphi (x)\rangle =|\psi (x)\rangle $ for any $x$.

Now we introduce the density operator as
\begin {equation}
\rho  = |\psi \rangle \langle \varphi | = |\psi \rangle \langle \psi |R,
\label {eq:32}
\end {equation}
which follows the evolution equation
\begin {equation}
  {\rm i} k^{-1}\frac {{\rm d}}{{\rm d} x}\rho  = [N, \rho ].
\label {eq:33}
\end {equation}

The matrix representation of $\rho $ is
given as follows:
\begin {equation}
  \rho  = \left [
  \begin {array}{cc}
-\alpha \beta ^*\; & |\alpha |^2 \\
-|\beta |^2    \; & \alpha ^*\beta
  \end {array}
  \right ].
\label {eq:34}
\end {equation}
It can be parametrized
as
\begin {equation}
\rho  = {\rm i}(s_0 I/2 + s_1 K_1 + s_2 K_2 + s_3 J_3),
\label {eq:35}
\end {equation}
where
\begin {eqnarray}
  s_0 &=& (\alpha ^*\beta  - \alpha \beta ^*)/{\rm i},\nonumber \\
  s_1 &=& \alpha ^*\beta  + \alpha \beta ^*,\nonumber \\
  s_2 &=& -|\alpha |^2 + |\beta |^2,\nonumber \\
  s_3 &=& |\alpha |^2 + |\beta |^2.
\label {eq:36}
\end {eqnarray}

Substitution of (\ref {eq:26}) and (\ref {eq:35})
into (\ref {eq:33}) gives the equation of motion for
$\bbox {s}=(s_1, s_2, s_3)$:
\begin {equation}
  k^{-1}\frac {{\rm d} \bbox {s}}{{\rm d} x} =
\bbox {\epsilon }\stackrel {\sim }{\times }\bbox {s},
\label {eq:37}
\end {equation}
which resembles to the Bloch equation, except for the use of
the SU(1,1) vector product.

We note that $s_0$ is a constant of motion and that the relation
\begin {equation}
  \bbox {s}\stackrel {\sim }{\cdot }\bbox {s} = -s_0^2
\label {eq:38}
\end {equation}
holds.

In the case of $s_0=0$,
the vector $\bbox {s}$ moves on the surface of a cone in the
$(s_1, s_2, s_3)$ space, as in Sec.\ \ref {sec:blochreal};
$s_0=0$ implies real amplitudes because $\alpha $ and $\beta $ share
the same phase, which can be eliminated.

In the case of $s_0\not =0$, $\bbox {s}$ moves on
the surface of a two-sheet hyperboloid, as shown in Fig.\ \ref {fig5}.
Without loss of generality, we normalize
the state vectors as $s_0=1$.

%===========================
\section {Geometrical Phase}
%===========================
\label {sec:geometricalphase}

In the case of real amplitudes, a density matrix represents two states.
On the other hand, in the case of complex amplitudes,
a density matrix $\rho $ represents
an infinite number of state vectors.
For example, the state
$|\psi '\rangle ={\rm e}^{-{\rm i}\phi  K_{s}}|\psi \rangle $,
$|\varphi '\rangle ={\rm e}^{{\rm i}\phi  K_{s}}|\varphi \rangle $
derived from $|\psi \rangle $ and $|\varphi \rangle $ with
$K_s=\bbox {s}\cdot \bbox {K}$
yield the same density matrix;
\begin {equation}
  \rho '=|\psi '\rangle \langle \varphi '|=
  {\rm e}^{-{\rm i}\phi  K_{s}}|\psi \rangle \langle \varphi |
  {\rm e}^{{\rm i}\phi  K_{s}} = \rho ,
\label {eq:40}
\end {equation}
since $K_s$ and $\rho =|\psi \rangle \langle \varphi |={\rm i}(s_0/2+K_{s})$
commute.
In other words, a density matrix or a Bloch vector
corresponds to an equivalence
class of states:
  $\left \{{\rm e}^{-{\rm i}\phi  K_{s}}|\psi \rangle \left |\right .
  -2\pi \le \phi <2\pi \right \}$,
or a ray.

Now we have a projective structure in which we can introduce the
Aharonov-Anandan connection \cite {aa}, thereby we can derive the dynamical
phase and the geometrical phase associated with a cycle of evolution.

The equivalence class introduced above can be represented more
conveniently by using a decomposition of the group element of
SL(2,R):
\begin {equation}
  g = {\rm e}^{-{\rm i}\mu  J_3} {\rm e}^{-{\rm i}\nu  K_1}
{\rm e}^{-{\rm i}\phi  J_3}.
\label {eq:41}
\end {equation}
An arbitrary state $|\psi \rangle $ can be obtained from
$|\psi _0\rangle =2^{-1/2}(1, {\rm i})^{\rm  T}$, which belongs to a ray
$\bbox {s}=(0,0,1)$ as
\begin {equation}
  |\psi \rangle =g|\psi _0\rangle ,\;
\langle \varphi |=\langle \varphi _0|g^{-1},
\label {eq:42}
\end {equation}
where $|\varphi _0\rangle =R^{-1}|\psi _0\rangle $.
The density matrix for this state is
\begin {equation}
  \rho =|\psi \rangle \langle \varphi |=
  {\rm e}^{-{\rm i}\mu  J_3} {\rm e}^{-{\rm i}\nu  K_1}|\psi _0\rangle
  \langle \varphi _0|{\rm e}^{{\rm i}\nu  K_1}{\rm e}^{{\rm i}\mu  J_3},
\label {eq:43}
\end {equation}
which does not contain the parameter $\phi $.

Now we see that
all the states in a ray can be derived by applying a group action
to the states $|\psi _{\phi }\rangle
={\rm e}^{-{\rm i}\phi  J_3}|\psi _0\rangle $,
whose real and imaginary
parts are given as
\begin {equation}
  |\psi _{\phi }'\rangle =\left [\begin {array}{c}\mathop {\rm cos}
\nolimits \phi /2
 \\ \mathop {\rm sin}\nolimits \phi /2 \end {array}\right ],
  \hskip 1em\relax
  |\psi _{\phi }''\rangle =\left [\begin {array}{c}-
\mathop {\rm sin}\nolimits \phi /2
 \\ \mathop {\rm cos}\nolimits \phi /2 \end {array}\right ].
\label {eq:44}
\end {equation}
Thus the element in a ray can be parametrized by $\phi $.

Reflecting the two-valuedness discussed in Sec.\ \ref {sec:twovalue},
the states with $\phi $ and $\phi +2\pi $ in a ray have opposite signs.

Let us consider a cyclic evolution in which a state
$|\psi (0)\rangle $ evolves and returns to a state
$|\psi (L)\rangle =\exp {({\rm i}\Phi )}|\psi (0)\rangle $, which
belongs to the same ray, i.e., $\rho (0)=\rho (L)$.
The phase difference $\Phi $
is composed of a dynamical phase $\phi _{\rm D}$ and a geometrical
phase $\phi _{\rm G}$.

The geometrical phase $\phi _{\rm G}$
or the Berry phase for this system is
given as \cite {kitano-yabuzaki,jordan}
\begin {equation}
  \phi _{\rm G}=\intop \nolimits _{\rm  S}s_3^{-1} {\rm d} s_1
\wedge {\rm d} s_2,
\label {eq:45}
\end {equation}
where S is the area enclosed by the path.
The integrand is an invariant two form under the group action.
It should be noted that this two form is different from that
for the surface curvature of
the hyperboloid, $(2s_3^2 -1 )^{-2}{\rm d} s_1\wedge {\rm d} s_2$, or
that for the surface area of
the hyperboloid, $(2 - s_3^{-3})^{1/2}{\rm d} s_1\wedge {\rm d} s_2$.
In the SU(2) case, all of the three quantities have the same form
incidentally.

%====================================
\section {Conclusions}
%====================================
\label {sec:conclusion}
In conclusion, we have explored geometrical features of 1D wave
propagation.
Underlying group of this problem is the (2+1)-dimensional Lorentz group.
The geometrical approach would be very helpful for intuitive understanding
of various wave phenomena.
It is also applicable to other systems which are
governed by the Lorentz dynamics.

Finally we would like to propose an experiment to observe the Berry phase
and other geometrical phenomena for the 1D wave propagation.
In Fig.\ \ref {fig6}, an optical system is presented.
We have a medium whose distribution of refractive index $\epsilon (x)$ can
be controlled externally.
It is possible to write a holographic grating in optical fibers or in some
nonlinear media.
The light propagated through the medium is reflected back with
perfect mirror M\@ .
Thus we can prepare a state corresponding to
$\bbox {s}(0)=(0,0,1)$ on the mirror side ($x=0$).
By adjusting the mirror position, we can ``tune'' the phase
$\phi _0$ of the initial state as
$|\psi (0)\rangle =(\mathop {\rm cos}\nolimits \phi _0/2,
\mathop {\rm sin}\nolimits \phi _0/2)^{\rm  T}$
and hence we can map out the complex $|\psi \rangle $ via
Eq.\ (\ref {eq:44}).
If we set up $\epsilon (x)$ so that the evolution be cyclic,
then the state on the other side ($x=L$) should be
$|\psi (L)\rangle =(\mathop {\rm cos}\nolimits \phi _L/2,
\mathop {\rm sin}\nolimits \phi _L/2)^{\rm  T}$.
For a cyclic evolution, $\Phi =\phi _L-\phi _0$ is independent of
$\phi _0$.
With the Mach-Zehnder interferometer
we can measure the difference $\Phi =\phi _{\rm G}+\phi _{\rm D}$.

\acknowledgments
The author is grateful to Professor H. Ogura for valuable
discussions.
This work is partly supported by the Ministry of Education, Science,
and Culture in Japan, under a Grant-in-Aide for Scientific Research.

\begin {figure}
\caption {
Bloch vector $\protect \bbox {s}$
(for real amplitudes) on the conic surface:
$s_1^2+s_2^2-s_3^2=0$
}
\label {fig1}
\end {figure}

\begin {figure}
\caption {
Waves with opposite polarities (real lines) have the
same phase relationship with the half-periodic
potential (dashed line).
}
\label {fig2}
\end {figure}

\begin {figure}
\caption {
In order to identify $(A,B)$ and $(-A,-B)$, which represent
physically the same state, we make a twofold cone
from the $(A,B)$ plane by making a cut from the origin and wrapping.
}
\label {fig3}
\end {figure}

\begin {figure}
\caption {
Interferometric detection of two-valuedness of wave amplitudes.
(a) Two trajectories starting off from $(A, B)=(1,0)$ at $x=0$
evolved by $\protect \bbox {\epsilon }=\pi (-2,1,3)/400$ (real line), and
by $-\protect \bbox {\epsilon }$ (dashed line),
respectively, and reach conjugate points at $x=100k^{-1}$.
(b) Corresponding wave forms.
}
\label {fig4}
\end {figure}

\begin {figure}
\caption {
Bloch vector $\protect \bbox {s}$ (for complex amplitudes) on the hyperboloid:
$s_1^2+s_2^2-s_3^2=-1$
}
\label {fig5}
\end {figure}

\begin {figure}
\caption {Proposed experimental setup for measuring the Berry phase
for a 1D wave system.}
\label {fig6}
\end {figure}

\end {document}
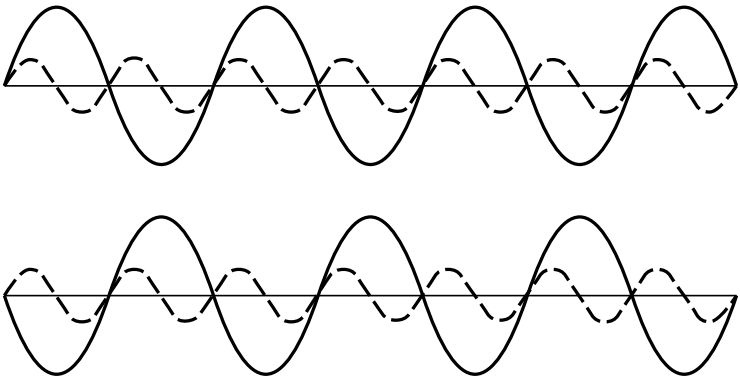
